\documentclass[fleqn,10pt]{wlscirep}
\title{Unsupervised decoding of long-term, naturalistic human neural recordings with automated video and audio annotations}
\author[1,2,3,4*]{Nancy X. R. Wang}
\author[4,5]{Jared D. Olson}
\author[2,4,6]{Jeffrey G. Ojemann}
\author[1,2,4]{Rajesh P. N. Rao}
\author[2,3,7]{Bingni W. Brunton}
\affil[1]{Dept. of Computer Science and Engineering, University of Washington, Seattle, 98195, USA}
\affil[2]{Institute for Neuroengineering, University of Washington, Seattle, 98195, USA}
\affil[3]{eScience Institute, University of Washington, Seattle, 98195, USA}
\affil[4]{Center for Sensorimotor Neural Engineering, University of Washington, Seattle, 98105, USA}
\affil[5]{Dept. of Rehabilitation Medicine, University of Washington, Seattle, 98195, USA}
\affil[6]{Dept. of Neurological Surgery, University of Washington, Seattle, 98195, USA}
\affil[7]{Dept. of Biology, University of Washington, Seattle, 98195, USA}

\affil[*]{wangnxr@uw.edu}

\begin{abstract}
Fully automated decoding of human activities and intentions from direct neural recordings is a tantalizing challenge in brain-computer interfacing. 
Most ongoing efforts have focused on training decoders on specific, stereotyped tasks in laboratory settings. 
Implementing brain-computer interfaces (BCIs) in natural settings requires adaptive strategies and scalable algorithms that require minimal supervision. 
Here we propose an unsupervised approach to decoding neural states from human brain recordings acquired in a naturalistic context.
We demonstrate our approach on continuous long-term electrocorticographic (ECoG) data recorded over many days from the brain surface of subjects in a hospital room, with simultaneous audio and video recordings. 
We first discovered clusters in high-dimensional ECoG recordings and then annotated coherent clusters using speech and movement labels extracted automatically from audio and video recordings.  
To our knowledge, this represents the first time techniques from computer vision and speech processing have been used for natural ECoG decoding. 
Our results show that our unsupervised approach can discover distinct behaviors from ECoG data, including moving, speaking and resting. 
We verify the accuracy of our approach by comparing to manual annotations. 
Projecting the discovered cluster centers back onto the brain, this technique opens the door to automated functional brain mapping in natural settings.

\end{abstract}
\begin{document}

\flushbottom
\maketitle
\thispagestyle{empty}

\section*{Introduction}

Much of our knowledge about neural computation in humans has been informed by data collected through carefully controlled experiments in laboratory conditions. 
Likewise, the success of Brain-Computer Interfaces (BCIs~\cite{Wolpaw2012,Rao2013})---controlling robotic prostheses and computer software via brain signals---has hinged on the availability of labeled data collected in controlled conditions.
Sources of behavioral and recording variations are actively avoided or minimized. 
However, it remains unclear to what extent these results generalize to naturalistic behavior.
It is known that neuronal responses may differ between experimental and freely behaving natural conditions~\cite{Vinje2000,Felsen2005,Jackson2007}.
Therefore, developing robust decoding algorithms that can cope with the challenges of naturalistic behavior is critical to deploying BCIs in real-life applications.

One strategy for decoding naturalistic brain data is to leverage external monitoring of behavior and the environment for interpreting neural activity.
Previous research that studied naturalistic human brain recordings, including brain surface electrocorticography (ECoG), have required ground truth labels~\cite{Derix2012,Pistohl2012,Ruescher2013}.
These labels were acquired by tedious and time-consuming manual labeling of video and audio.
In addition to being laborious, manual labeling is prone to human errors from factors such as loss of attention and fatigue~\cite{Hill2012}.
This problem is exacerbated by very long recordings, when patients are monitored continuously for several days or long. 
Obtaining labeled data and training algorithms extensively are difficult or even intractable in rapidly changing, naturalistic environments.

\begin{figure}[t]
\centering
\includegraphics[scale=0.4]{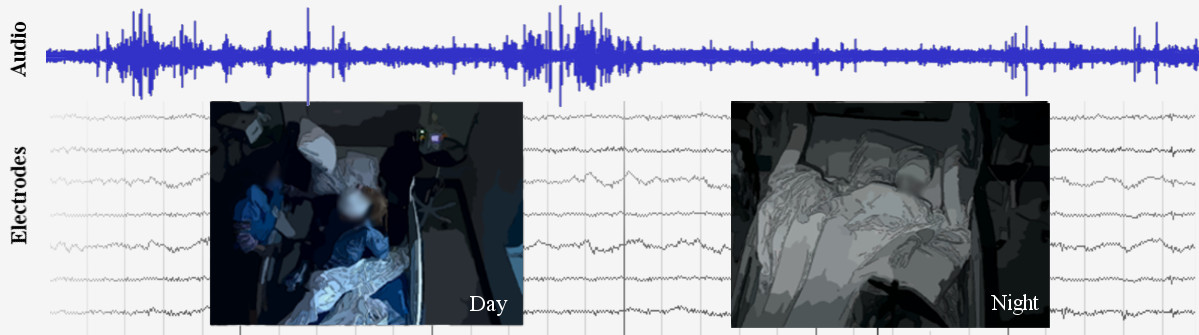}
\caption{An excerpt from the data set, which includes video, audio, and intracranial brain activity (ECoG) continuously recorded for at least one week for six subjects. ECoG recordings from a small subset of the electrodes are shown, along with the simultaneously recorded audio signals in blue. A typical patient has around 100 electrodes. Overlaid are screen shots of the video, which is centered on the patient; on the left is a daytime video of the patient eating, and on right is a nighttime infrared video of the patient sleeping.  For patient privacy, faces have been blurred.}
\label{fig:sample_frames}
\end{figure}

In this article, we describe our use of video and audio recordings in conjunction with ECoG data to decode human behaviour in a completely unsupervised manner. 
Fig.~\ref{fig:sample_frames} illustrates components of the data used in our approach. 
The data consists of six subjects monitored continuously over at least one week after electrode array implantation; each subject had approximately 100 intracranial ECoG electrodes with wide coverage of cortical areas.
Importantly, subjects being monitored had no instructions to perform specific tasks; they were undergoing presurgical epilepsy monitoring and behaved as they wished inside their hospital room.
Instead of relying on manual labels, we used computer vision, speech processing, and machine learning techniques to automatically determine the ground truth labels for the subjects' activities. 
These labels were used to annotate patterns of neural activity discovered using unsupervised clustering on power spectral features of the ECoG data. 
We demonstrate that this approach can identify salient behavioral categories in the ECoG data, such as movement, speech and rest.  
Decoding accuracy was verified by comparing the automatically discovered labels against manual labels of behavior in a small subset of the data. 
Further, projecting the annotated ECoG clusters to electrodes on the brain revealed spatial and power spectral patterns of cortical activation consistent with those characterized during controlled experiments. 
These results suggest that our unsupervised approach may offer a reliable and scalable way to map functional brain areas in natural settings and enable the deployment of BCI in real-life applications.

\subsection*{Background and Related Work}
Intracranial electrocorticography (ECoG) as a technique for observing human neural activity is particularly attractive. 
Its spatial and temporal resolution offers measurements of temporal dynamics inaccessible by functional magnetic resonance imaging (fMRI) and spatial resolution unavailable to extracranial electroencephalography (EEG).
Cortical surface ECoG is accomplished less invasively than with penetrating electrodes~\cite{Moran2010,Williams2007} and has much greater signal-to-noise ratio than entirely non-invasive techniques such as EEG~\cite{Lal,Ball2009}.

Efforts to decode neural activity are typically accomplished by training algorithms on tightly controlled experimental data with repeated trials.
Much progress has been made to decode arm trajectories~\cite{Nakanishi2013,Wang2012,Wang2013} and finger movements~\cite{Miller2009,Wang2010}, to control robotic arms~\cite{McMullen2014,Yanagisawa2011,Fifer2014}, and to construct ECoG  BCIs~\cite{Wang2013a,Leuthardt2011,Leuthardt2006,Miller2010,Schalk2008,Vansteensel2010}.
Speech detection and decoding from ECoG has been studied at the level of voice activity~\cite{Kanas2014b}, phoneme~\cite{Blakely2008,Leuthardt2011,Kanas2014a,Mugler2014}, vowels and consonants~\cite{Pei2011}, whole words~\cite{Kellis2010}, and sentences~\cite{Zhang2012}. 
Accurate speech reconstruction has also been shown to be possible~\cite{Herff2015}.

The concept of decoding naturalistic brain recordings is related to passive BCIs, a term used to describe BCI systems that decode arbitrary brain activity that are not necessarily under volitional control~\cite{Zander2011}. Our system, which falls within the class of passive BCIs, may also be considered a type of hybrid BCI combining electrophysiological recordings with other signals~\cite{Muller2015}.
However, past approaches in this domain have not focused on combining alternative monitoring modalities such as video and audio in order to decode natural ECoG signals. 

The lack of ground-truth data makes decoding naturalistic neural recordings difficult.
Supplementing neural recordings with additional modes of observation, such as video and audio, can make the decoding more feasible. 
Previous studies exploring this idea have decoded natural speech~\cite{Derix2012,Derix2014,Dastjerdi2013} and natural motions of grasping~\cite{Pistohl2012,Ruescher2013}; however, these studies relied on laborious manual annotations.
Entirely unsupervised approaches to decoding have previously targeted sleep stages~\cite{Langkvist2012} and seizures~\cite{Pluta2014} rather than long-term natural ECoG recordings. 

Our approach to circumvent the need for manually annotated behavioral labels exploits automated techniques developed in computer vision and speech processing. 
Both of these fields have seen tremendous growth in recent years with increasing processor power and advances in methodology~\cite{Jordan2015,Huang2014}.
Computer vision techniques have been developed for a variety of tasks including automated movement estimation~\cite{Wang2015,Poppe2007}, pose recognition~\cite{Toshev2014}, object recognition~\cite{Girshick2014,Erhan2014}, and activity classification~\cite{Karpathy2014,Ryoo2013}.
In some cases, computer vision techniques have matched or surpassed single-human performance in recognizing arbitrary objects~\cite{He2015}.
Voice activity detection has been well studied in speech processing~\cite{Ramrez2004}. In this work, we leverage and combine techniques from these rapidly advancing fields to automate and enhance the decoding of naturalistic human neural recordings.

\section*{Results}

\begin{figure}[t]
\centering
\includegraphics[width=0.9\linewidth]{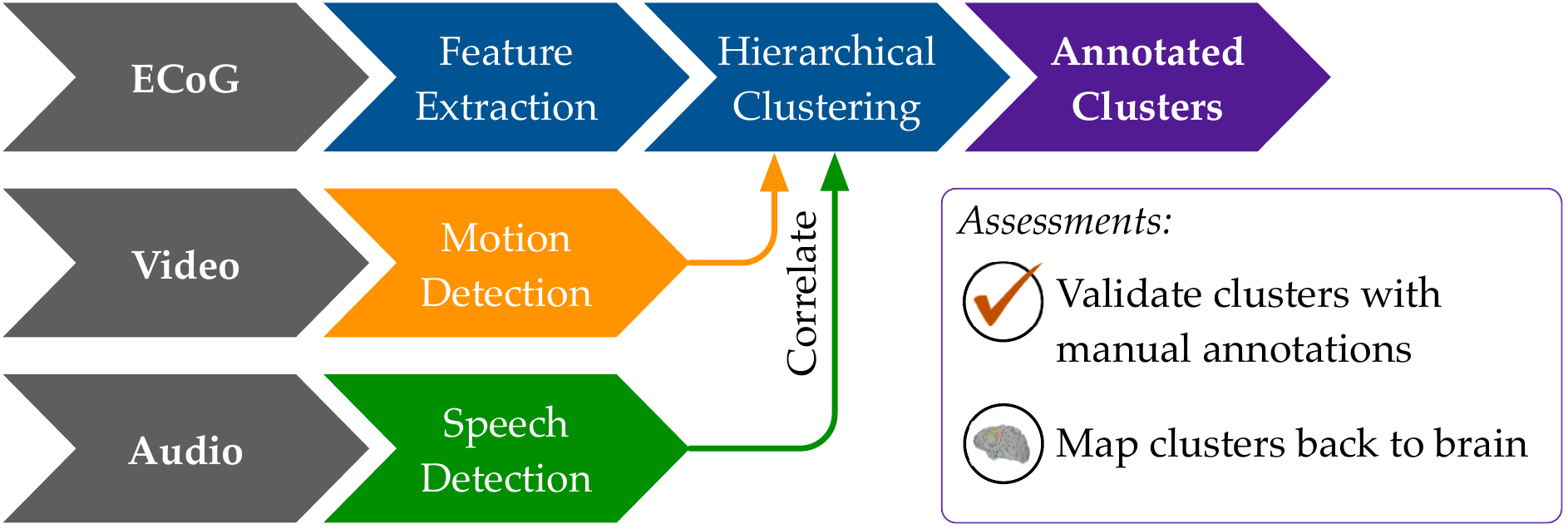}
\caption{An overview of our methods to discover neural decoders by automated clustering and cluster annotations.
Briefly, the ECoG recordings was broken into short, non-overlapping windows of 2-seconds. Power spectral features were extracted for each electrode, all electrodes' features were stacked, and the feature space was reduced to the first 50 principal component dimensions. Hierarchical k-means clustering was performed on these 50-dimensional data, and annotation was done by correlations in timing with automated detection of motion and speech levels (see Fig.~\ref{fig:hierarchy}). The resultant annotated clusters were validated against manual annotations; cluster centroids mapped to the brain visualize the automatically detected neural patterns. }
\label{fig:pipeline}
\end{figure}

Our general approach to unsupervised decoding of large, long-term human neural recordings is to combine hierarchical clustering of high-dimensional ECoG data with annotations informed by automated video and audio analysis, as illustrated in Fig.~\ref{fig:pipeline} (further details in the Methods section). Briefly, hierarchical k-means clustering was performed on power spectral features of the ECoG recordings.
These clusters are coherent patterns discovered in the neural recordings; video and audio monitoring data was used to interpret these patterns and match them to behaviorally salient categories such as movement, speech and rest.
Here we describe results of our analysis on six subjects where we used automated audio and video analysis to annotate clusters of neural activity.
The accuracy of the unsupervised decoding method was quantified by comparison to manual labels, and the annotated clusters were mapped back to the brain to enable neurologically relevant interpretations.

\subsection*{Actograms: automated motion and speech detection}

\begin{figure}[t]
\centering
\includegraphics[scale=0.7]{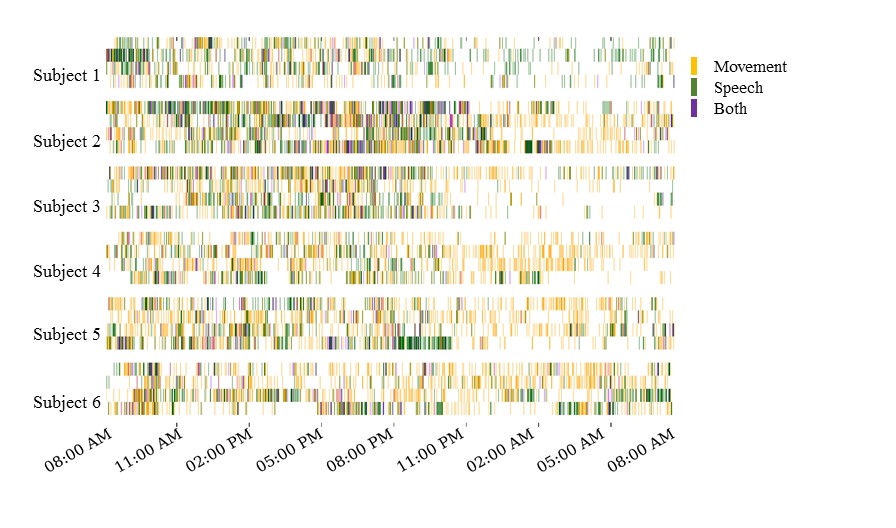}
\vspace{-5pt}
\caption{Daily actograms for all subjects. Each row shows one day of activity profiles summarized by automated speech and motion recognition algorithms. Days 3--6 post surgical implantation were analyzed. For purposes of this visualization, the activity levels were binned to one-minute resolution. Movement and speech levels were highly correlated on most days, and these were concentrated to the active hours between 8:00 AM and 11:00PM. }
\label{fig:actogram}
\end{figure}

The automated motion and speech detection methods quantified movement and speech levels from the video and audio recordings, respectively.
Fig.~\ref{fig:actogram} shows daily ``actograms'' for all six subjects.
Movement levels were quantified by analyzing magnitude of changes at feature points in successive frames of the video.
Speech levels were quantified by computing the power in the audio signal in the human speech range.

As expected, Fig.~\ref{fig:actogram} shows that subjects were most active during waking hours, generally between 8:00AM and 11:00PM.
Also, movement and speech levels are often highly correlated, as the subjects were often moving and speaking at the same time during waking hours. 
During night time hours, although subjects were generally less active, many instances of movement and speech can still be seen in Fig.~\ref{fig:actogram} as the subjects either shifted in their sleep or were visited by hospital staff during the night.

Our automated motion and speech detection algorithms were able to perform with reasonable accuracy when compared to manual annotation of movement and speech.
Over all subject days, movement detection was 74\% accurate (range of 68\% to 90\%), while speech detection was 75\% accurate (range of 67\% to 83\%).

\subsection*{Unsupervised decoding of ECoG activity}

Unsupervised decoding of neural recordings was performed by hierarchical clustering of power spectral features of the multi-electrode ECoG recordings.
Because the subjects' behavior on each day varied widely, both across days for the same subject and across subjects (see actograms in Fig.~\ref{fig:actogram}), we were agnostic to which specific frequency ranges contained meaningful information and considered all power in frequency bins between 1 and 53 Hz (see Supplemental Information for results considering higher frequency bands).
Further, data from each subject day was analyzed separately.
Clusters identified by hierarchical k-means clustering were annotated using information from the external monitoring by video and audio.
The hierarchical k-means clustering implementation is detailed in the Methods section. Following a tree structure, successive levels of clustering contained larger numbers of clusters (Fig.~\ref{fig:hierarchy}).

\begin{figure}[t]
\includegraphics[scale=0.44]{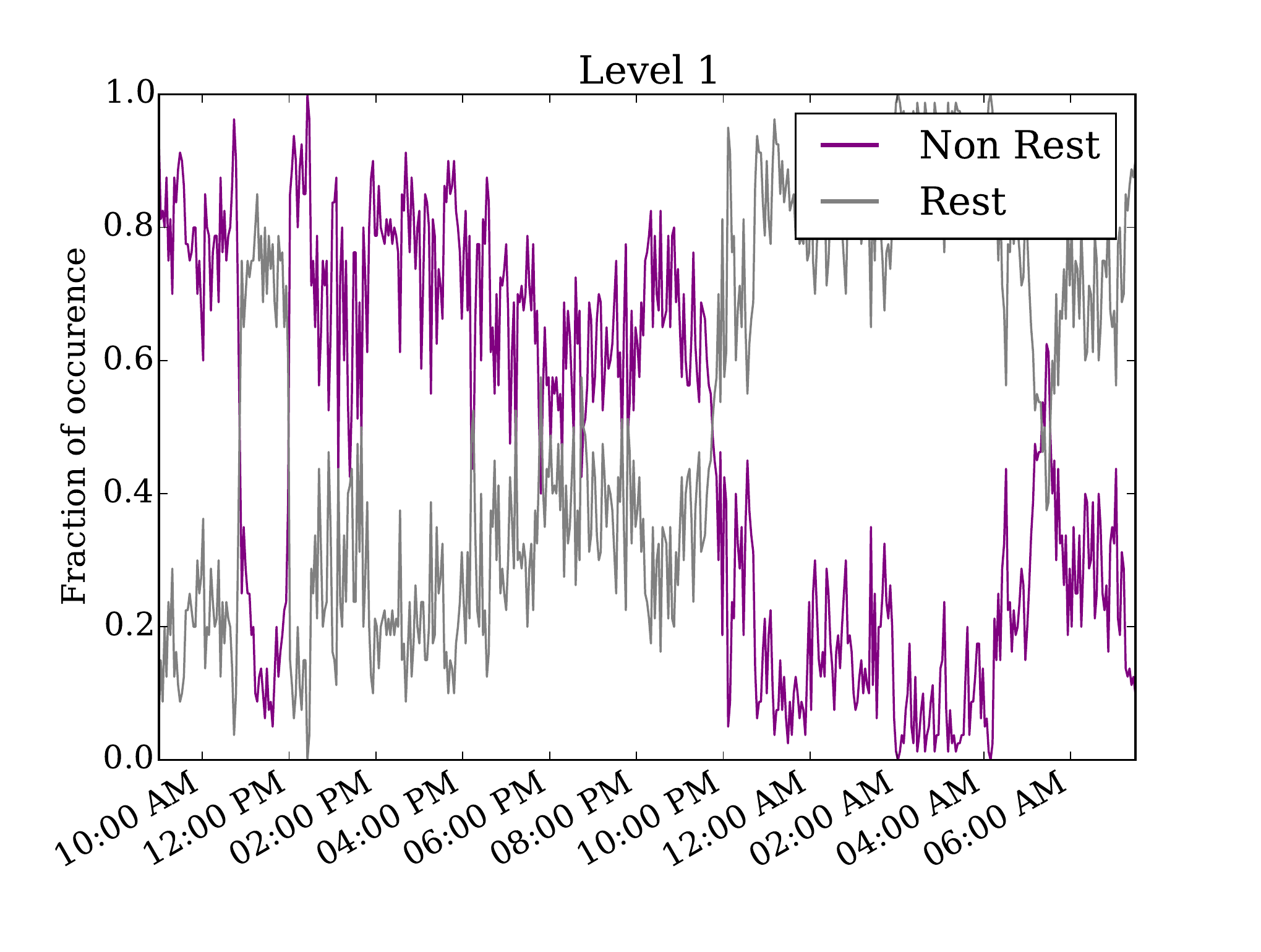}
\includegraphics[scale=0.44]{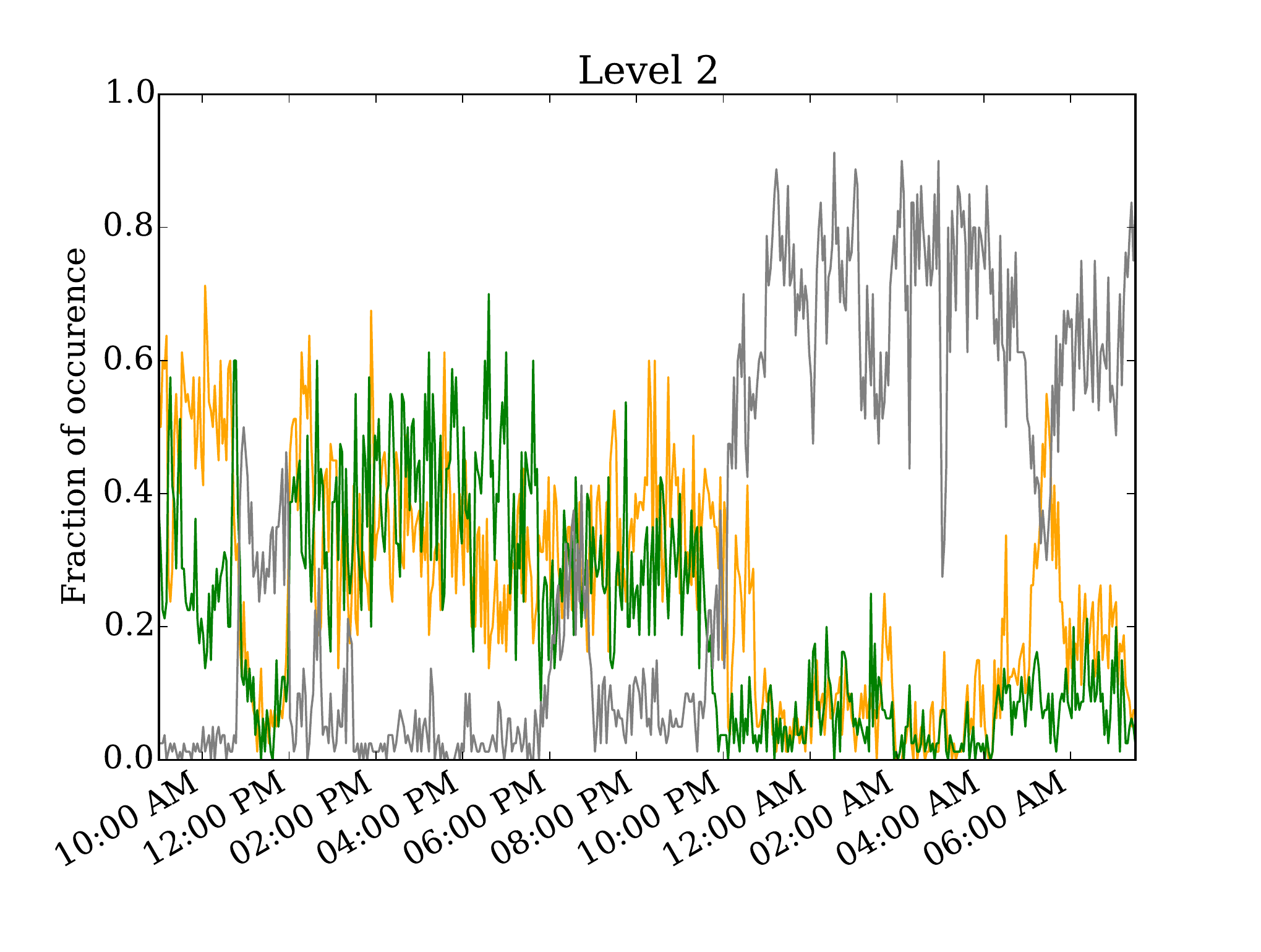}
\includegraphics[scale=0.44]{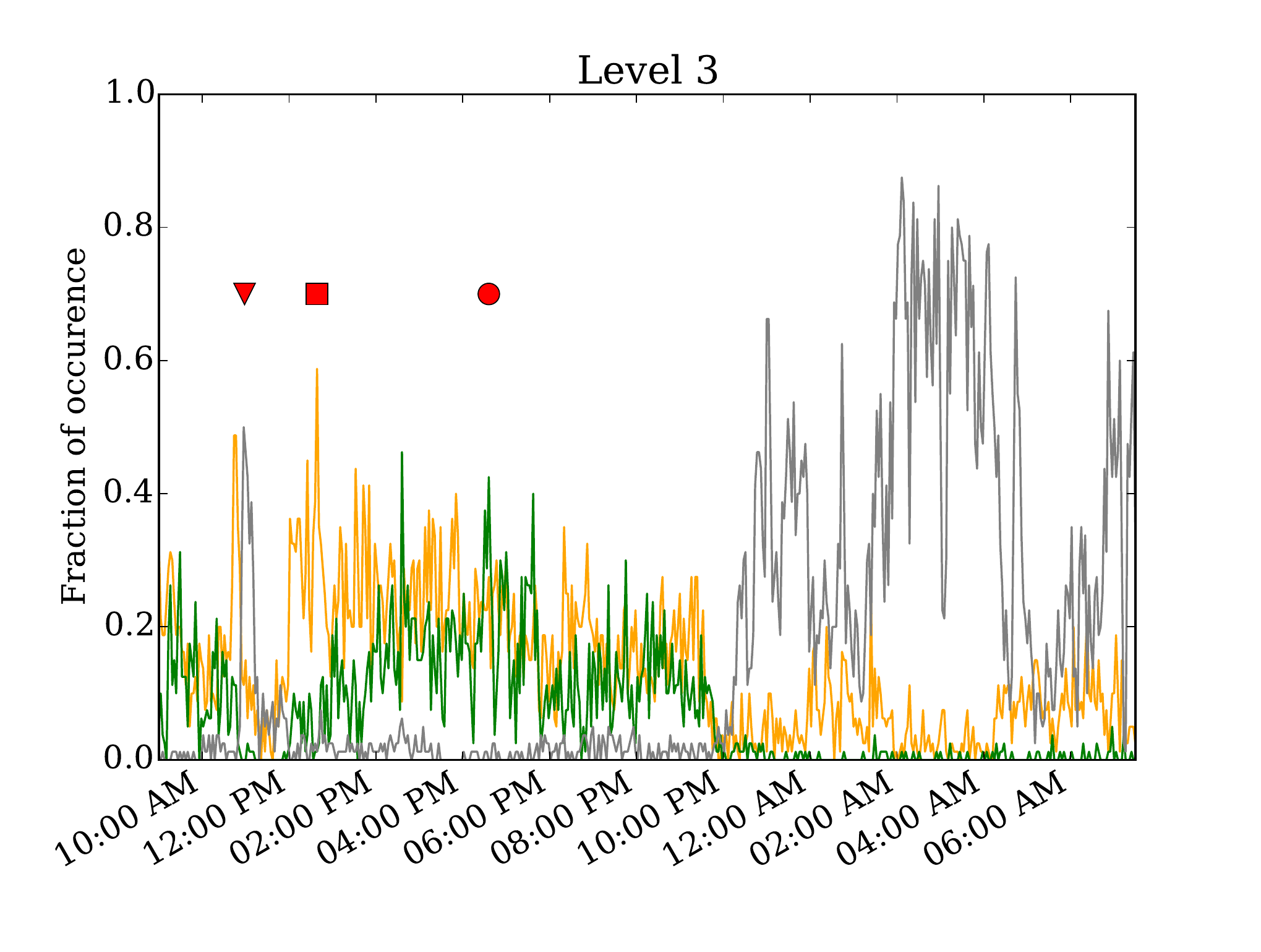}
\includegraphics[scale=0.44]{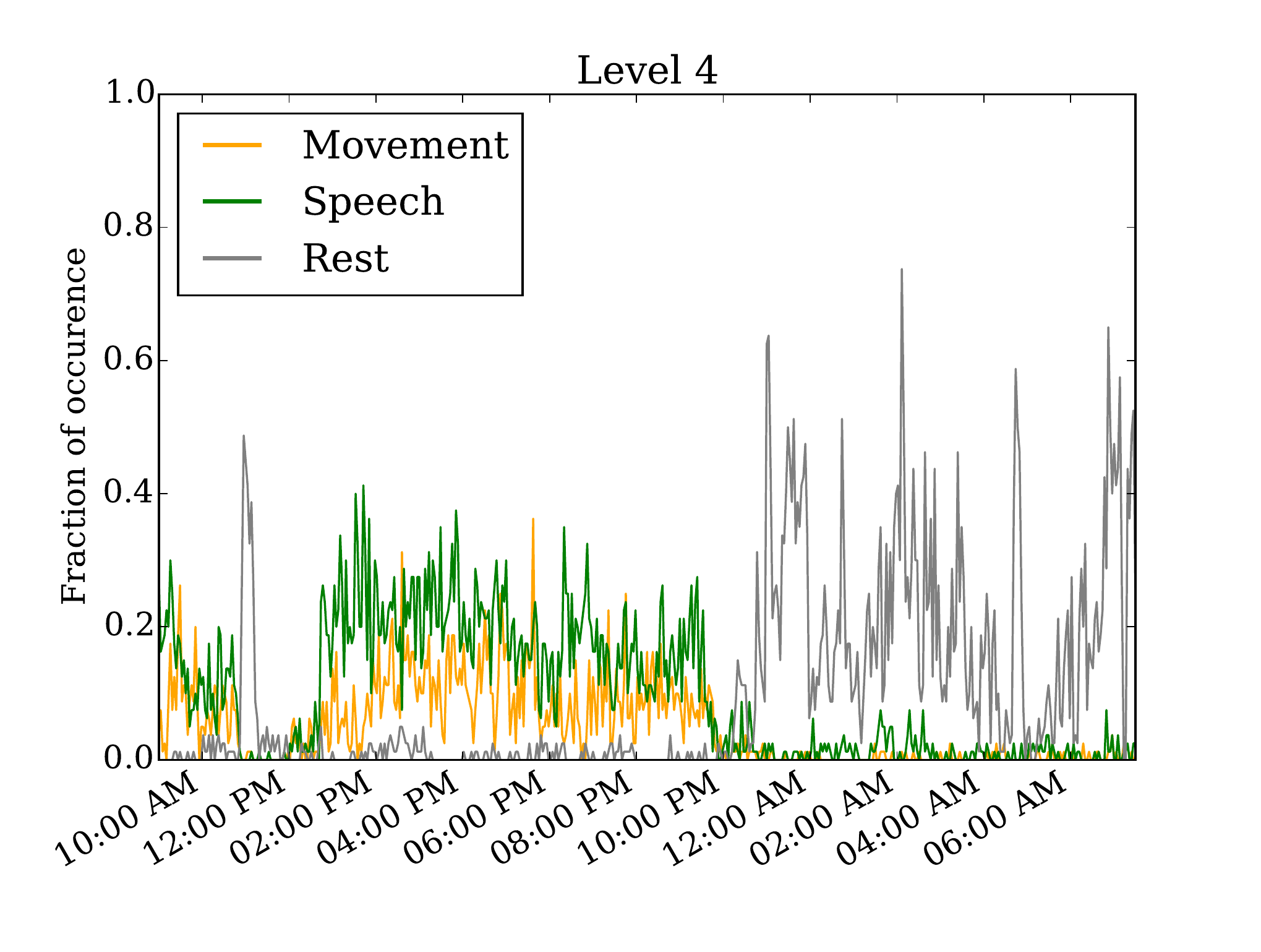}
\caption{Annotated clustering results of one subject day (Subject 6 on day 6 post implant) from hierarchical level 1 to level 4. The vertical axis represents the fraction of time the neural recording is categorized to each annotated cluster. The triangle marks when the subject takes a nap (Video~S2), the square marks when the subject is seen to move without speech (Video~S3), and the circle marks when the subject spoke more than moved (Video~S4). For visualization, the 24-hour day was binned to every 160 seconds.}
\label{fig:clusters_annotated}
\end{figure}

Fig.~\ref{fig:clusters_annotated} shows results of the annotated clusters for one subject day (Subject 6 on day 6 post implantation) at clustering levels 1--4 as a function of time of day.
At level 1, it is clear that rest is separable from non-rest, and the switch in the dominant cluster occurred around 10:00PM.
We presume the timing of the switch to correspond to when the subject falls asleep, as is corroborated by the video monitoring.
when the subject is presumed to have fallen asleep as evident in the video monitoring.
Video~S1 shows an example of the infrared video acquired during night time.
The subject is in a consolidated period of rest between 10:00PM and 9:00AM the following day.
Interestingly, for a duration of approximately one hour starting at around 11:00AM, the rest cluster dominated the labels (see also red triangle at level 3).
This period corresponds to the subject taking a nap (Video~S2).

Starting at level 2, the non-rest behavior separates into movement and speech clusters. 
These two clusters are generally highly correlated, as moving and talking often co-occur, especially as the movement quantification can detect mouth or face movement.
We point out several interesting instances labeled at level 3.
First, the inverted triangle points to a period around 11:00AM annotated as rest, when the subject rested during a nap (Video~S2).
Second, the rectangle marks a period around 1:00PM, annotated as predominantly movement but not speech, when the subject shifted around in their bed but did not engage in conversation (Video~S3).
Third, the circle marks a period around 5:00PM when the subject engaged in conversation (Video~S4); this period was labeled as both movement and speech.
As described in the validation analysis in the following section, the accuracy of the automated annotations does not change substantially between levels 3--4 across all subject days (Fig.~\ref{fig:percentile_scattergories} and Fig.~S3).

\subsection*{Validation of automated neural decoding by comparison with manual annotations}

\begin{figure}[t]
\centering
\includegraphics[scale=0.59]{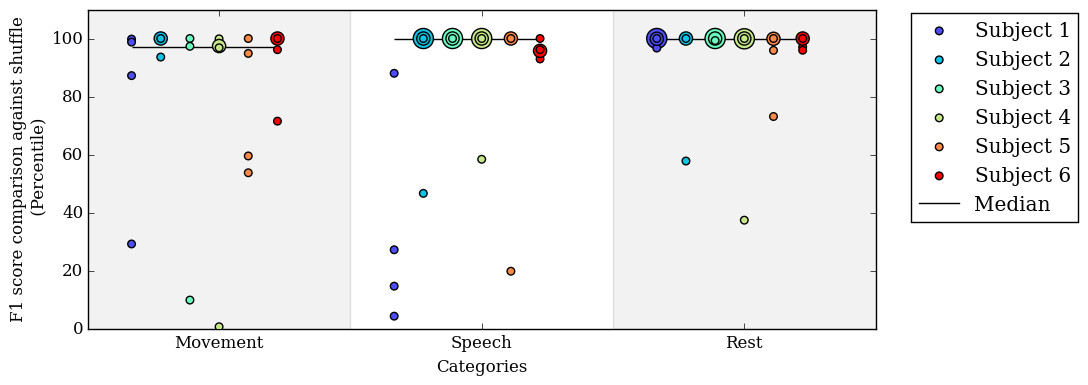}
\caption{Percentile of the F1 score of our algorithm at level 3 compared to F1 scores from randomly shuffled manual labels. Each colored dot corresponds to one day for a subject in each behavioral category.}
\label{fig:percentile_scattergories}
\end{figure}

The automated neural decoding was assessed by comparison with behaviors labeled manually.
Manual labels of the video and audio were supplied by two human annotators, who labeled a variety of salient behaviors for at least 40 total minutes (or approximately 3\%) of video and audio recordings for  each subject day.
The labels were acquired for 2-minute segments of data distributed randomly throughout the 24-hour day.

The entirely automated neural decoding performed very well in the validation for all subjects on the categories of movement, speech and rest.
Table~\ref{table:accuracy} summarizes the accuracy of the annotated clusters averaged over the 4 days analyzed for each subject, comparing the automated labels to manual labels during the labeled portions of each day.
In addition to computing the accuracy, we also computed the F1 scores of the automated decoding using manual labels as ground truth for each day; the F1 score is a weighted average of precision and recall (Table~S1).

To assess the significance of the automated labels' accuracy, we compared the F1 scores on each day to F1 scores of randomly shuffled labels.
The shuffled labels preserved the relative occurrence of labels and gave an unbiased estimate of chance performance.
Fig.~\ref{fig:percentile_scattergories} shows the percentile of the true F1 scores within the randomly shuffled F1 scores at hierarchical clustering level 3.
For each category of movement, speech and rest, the median percentile of the true F1 scores are at or near the 99th percentile; our automatically labeled clusters performed significantly better than chance on most subject days.
F1 score percentiles for clustering levels 2 and 4 are shown in Figs~S2 and S3.
We also repeated the analysis considering spectral frequencies up to 105 Hz, which does not substantially change the performance of the automated decoder (Table~S2 and Fig.~S4)

\begin{table}
\centering
\begin{tabular}{lrrr}
\begin{tabular}{lrrr}
\hline
    &   Movement &   Speech &   Rest \\
\hline
 Subject 1 &      59.06 &    54.24 &  64.98 \\
 Subject 2 &      62.88 &    62.20 &  64.47 \\
 Subject 3 &      61.22 &    62.94 &  64.75 \\
 Subject 4 &      58.57 &    60.43 &  65.88 \\
 Subject 5 &      55.61 &    60.01 &  58.09 \\
 Subject 6 &      70.08 &    69.43 &  57.60  \\
\hline
\end{tabular}
\end{tabular}
\caption{Percent accuracy as assessed by comparison of level 3 automated cluster annotation to manual annotations  averaged over all 4 days for each subject.} \label{table:accuracy}
\end{table}

\subsection*{Neural correlates of behavior as discovered by unsupervised clustering}

\begin{figure}[t]
\includegraphics[scale=0.5]{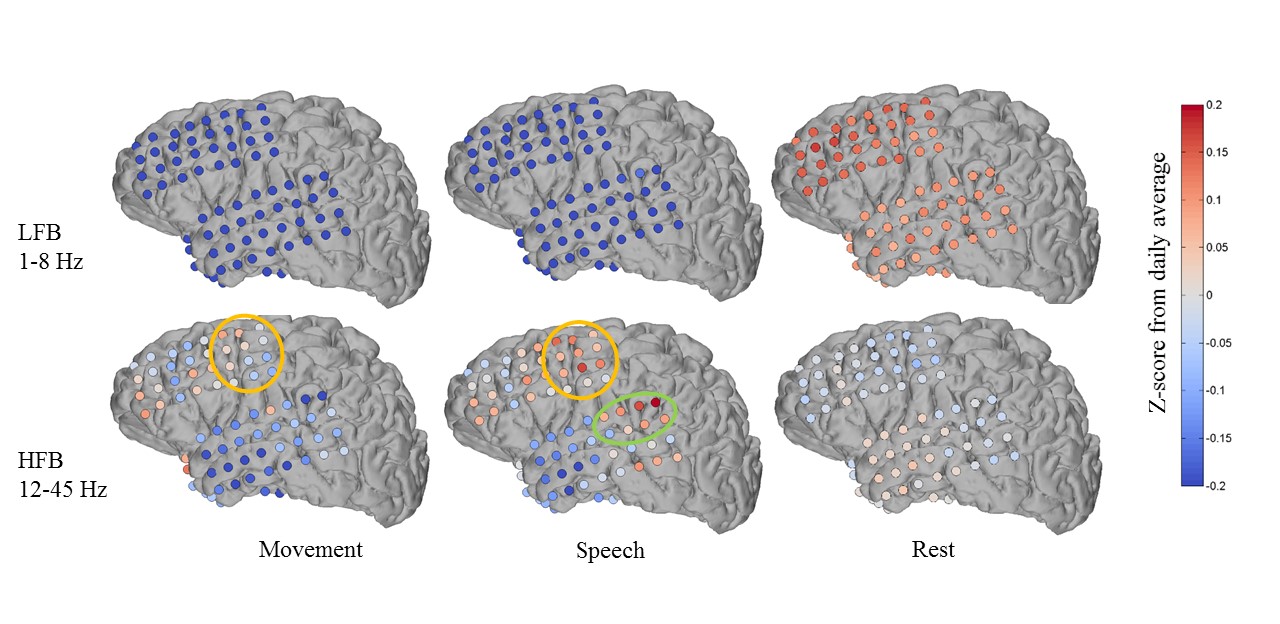}
\caption{Features discovered by automated brain decoding at two different frequency bands are consistent with known functions of cortical areas.
Shown for one subject day (Subject 1 on day 6 post implant), the centroids of the movement, speech and rest clusters were back projected to brain-electrode space, and then separately averaged over a low frequency band (LFB, 1--8Hz) and a high frequency band (HFB, 12--45Hz). The orange and green circles mark the approximate extent of locations typically considered to be sensorimotor and auditory regions, respectively. The colormap indicates the Z-Score of the power levels as compared to the daily average.}
\label{fig:backprojection}
\end{figure}

Another way to assess the neural decoder discovered through clustering and automated annotation is to examine the neural patterns identified in this unsupervised approach.
We mapped these patterns by projecting the centroids of annotated clusters back to feature space.
Next, the feature space in electrode coordinates on the brain were averaged within frequency bands, including those typically of interest to studies of human ECoG.

Fig.~\ref{fig:backprojection} shows an example of one subject day's annotated cluster centroids shown as deviations from the daily average in a low frequency band (LFB, 1--8Hz) and a high frequency band (HFB, 12--45Hz).
The LFB was chosen to include activity in the delta and theta range, while the HFB includes beta and low gamma activity.
The accuracy of automated decoding on this subject day (Subject 1, day 6 post implant) was 0.56, 0.69 and 0.63 for movement, speech and rest, respectively.
In the LFB, there was generalized decrease in power across all recording electrodes during movement and speech, accompanied by a corresponding relative increase in power during rest.
In contrast, in the HFB during movement and speech, we observe more spatially specific increase in power that is localized to motor areas (orange circle in Fig.~\ref{fig:backprojection}).
There is some overlap in electrodes showing increased HFB power during movement and speech, which may be due to activation of motor areas to produce speech.
In addition, during speech but not during movement, there is a localized increase in HFB power at associated auditory region (green circle in Fig.~\ref{fig:backprojection}).

These features are largely consistent with known functions of human cortical areas and ECoG phenomena, as well as the existing ECoG literature on motor activation~\cite{Miller2007,Miller2009} and speech mapping~\cite{Potes2014,Chang2010}.
We must note that these patterns of frequency band-specific changes in power for different behavioral categories were discovered in an entirely unsupervised approach, using continuously acquired naturalistic data, and without the luxury of subtraction of baseline activation immediately before or after movement.
It is important to keep in mind that previous studies typically define rest as the time just before an action, whereas we compare to daily averages as well as to sleep. 
During non rapid eye movement sleep, the theta and delta bands tend to have high power~\cite{Cajochen1999}, a factor that distinguishes our results from those obtained from more controlled experiments.  
We observed qualitatively similar patterns across the four (4) subjects where anatomic reconstruction of the electrode arrays were available (Figs.~S6--S8).

\section*{Discussion}
Our results represent, to our knowledge, the first demonstration of automated clustering and labeling of human behavior from brain recordings in a naturalistic setting; we achieved annotation without manual labels by leveraging techniques from computer vision and speech processing. 
Our unsupervised approach discovers clusters for behaviors such as moving, speaking and resting from ECoG data. 
The discovered cluster labels were verified by comparison to manual labels for a subset of the data. 
We also demonstrate that projecting the cluster centers back onto the brain provides an avenue for automated functional brain mapping in natural settings. 

Our goal was to develop an approach to decode human brain recordings by embracing the richness and variability of complex naturalistic behavior, while avoiding tedious manual annotation of data and fine tuning of parameters.
Our current approach has a number of limitations which can be addressed by improving both the available information streams and the algorithmic processing. 
One limitation of our movement detection algorithm is lack of specificity to the subject when other people enter the frame of the camera.
This is particularly challenging when another person overlaps with the subject, for example, when a nurse examines the patient.
We are exploring the potential of better subject segmentation using a depth camera. 
The depth stream information will also allow us to perform much more detailed pose recognition, including obtaining specific movement information from isolated body parts.

A second limitation is our inability to identify the speaker in speech detection. 
Speech levels include the subject speaking, the subject listening to another person speaking in the room, and the subject listening to the TV or another electronic audio source.
We expect that by placing an additional microphone in the room and using algorithms to distinguish speaker voices, it may be possible to more accurately localize the speaker and speech sources.

The temporal aspect of high-dimensional, long-term ECoG data may be better exploited to improve the clusters discovered by unsupervised pattern recognition techniques.
For instance, dimensionality reduction by dynamic mode decomposition (DMD~\cite{Brunton2015}) may be able to identify spatio-temporal patterns when repeated trials are not available.
Phase synchrony and phase coupling may also serve as important neural correlates of behavior~\cite{Mercier2015}.

Overall, these results demonstrate that our method has the practicality and accuracy to passively monitor the brain and decode its state during a variety of activities.
In our results, we see some variation in performance and cluster maps across days for the same subject.
This variance may be due to changes in brain activity as the patient recovers from surgery, or it may represent natural variation from day to day. 

Functional brain mapping acquired by analyzing neural recordings outside instructed tasks has direct relevance to how an individual brain functions in natural conditions.
For instance, neural correlates of a subject repeating a series of specific actions may differ from the full range of neural signatures associated with movements in general.
Previous attempts to do more ``ethological'' mapping based on non-cued activities have identified motor~\cite{Vansteensel2013,Breshears2012} and speech~\cite{Derix2012, Derix2014} related areas.
Our approach to ethological functional brain mapping explores the analysis of task-free, naturalistic neural data augmented by information from external monitoring, which enables us to perform the automated analysis at a much larger scale with long-term data.

We envision our automated passive monitoring and decoding approach with video and audio as a possible strategy to adjust for natural variation and drift in brain activity without the necessity to retrain decoders explicitly.
Such an approach may enable deployment of long-term BCI systems, including clinical and consumer applications.
More generally, we believe the exploration of large, unstructured, naturalistic neural recordings will improve our understanding of the human brain in action.

\section*{Methods}

\subsection*{Subjects and recording}
All six subjects had a macro-grid and one or more strips of electrocorticography (ECoG) electrodes implanted subdurally for presurgical clinical epilepsy monitoring at Harborview Medical Center. 
The study was approved by University of Washington's Institutional Review Board's human subject division; all subjects gave their informed consent.

Electrode grids were constructed of 3-mm-diameter platinum pads spaced at 1 cm center-to-center and embedded in silastic (AdTech). 
Electrode placement and duration of each patient's recording were determined solely based on clinical needs. 
The number of electrodes ranged from 82 to 106, arranged as grids of 8$\times$8, 8$\times$4, 8$\times$2 or strips of 1$\times$4, 1$\times$6, 1$\times$8. 
Figs.~S5--S9 show the electrode placements of each subject. 
ECoG was acquired at a sampling rate of 999Hz. 
All patients had between six and fourteen days of continuous monitoring with video, audio, and ECoG recordings. 
During days 1 and 2, patients were generally recovering from surgery and spent most of their time sleeping; in this study, days 3 to 6 post implant were analyzed from each subject.

\subsection*{Video and audio recordings}
Video and audio were recorded simultaneously with the ECoG signals and continuously throughout the subjects' clinical monitoring. 
The video was recorded at 30 frames per second at a resolution of 640$\times$480 pixels. 
Generally, video was centered on the subject with family members or staff occasionally entering the scene. 
The camera was also sometimes adjusted throughout the day by hospital staff; for instance, the camera may be centered away during bed pan changes and returned to the patient afterwards. 
Videos~S1--S4 show examples of the video at a few different times of one day.
The audio signal was recorded at 48 KHz in stereo. 
The subject's conversations with people in the hospital room, including people not visible by video monitoring, can be clearly heard, as well as sound from the television or a music player.
Some subjects listened to audio using headphones, which were not available to our audio monitoring system.
For patient privacy, because voices can be identifiable, we do not make examples of the audio data available in the supplemental materials.

\subsection*{Manual annotation of video and audio}
To generate a set of ground-truth labels so that we may assess the performance of our automated algorithms, we performed manual annotation of behavior aided by ANVIL ~\cite{kipp2012} on a small subset of the external monitoring data. 
Two students were responsible for the annotations, and at least 40 minutes (or 2.78\%) of each subject day's recording was manually labeled for a variety of salient behaviors, including the broad categories of movement, speech and rest.
Manual labeling was done for 2-minute segments of video and audio, distributed randomly throughout each 24-hour day.
For patient privacy, some small parts of the video (e.g., during bed pan changes) were excluded from manual labeling.
These periods were very brief and should not introduce a generalized bias in manual labels.
On average, manual labeling of 1 min of monitoring data was accomplished in approximately 5 minutes.
At least 10 minutes of each subject day were labeled by both students; agreement between the two labelers was 92.0\%, and Cohen's Kappa value for inter-rater agreement is 0.82.

\subsection*{Automated movement and speech detection}

For automated video analysis, we first detected salient features for each frame using Speeded Up Robust Features (SURF), which detects and encodes interesting feature points throughout the frame. 
The amount of motion in each frame was determined by matching the magnitude of change in these feature points across successive frames.
Since the subject was the only person in the frame a majority of the time, we are able to determine the subject's approximate movement levels.
This approach detected gross motor movements of the arms, torso and head, as well as some finer movements of the face and mouth during speaking. 
To detect speech, we measured the power of human speech frequency levels (100--3500Hz) from audio data. 

We assessed the performance of the automated algorithms by comparison to manual annotations.
The manual annotations for each behavior were binary (i.e., either the behavior was present or not in a time window) whereas the automated speech and movement levels were analog values.
Therefore, the agreement was computed after applying a threshold to the automated movement and speech levels.

\subsection*{ECoG preprocessing and feature extraction}
All ECoG recording was bandpassed filtered between 0.1 and 160Hz to reduced noise. 
The filtered signal was then converted into a set of power spectral features using short-time Fourier transform using non-overlapping two-second windows.
Each 24-hour recording was thus separated into 43200 samples in time.

Because subjects engaged in a variety of activities throughout the day, there is no particular frequency band that would be solely useful for clustering. 
We considered power at a range of frequencies between 1--52 Hz for each electrode, binning every 1.5 Hz of power for a total of 35 features per electrode per two-second window.
At 82 to 106 electrodes per subject, this process resulted in 2870 to 3710 features for each two-second window of recording. 
To normalize the data, we transformed the binned powers levels at each frequency bin for each channel by computing the Z-Score.
The dimensionality of the feature space was then reduced with principal component analysis (PCA), and the cumulative fraction of variance explained as a function of the number of PCA modes for each subject day is shown in Fig.~S1.
These spectra were highly variable, both within and between subjects.
For purposes of unsupervised clustering, we truncated all feature space to the first 50 PC's.
The first 50 PC's generally accounted for at least 40\% of the variance in daily power spectral features space.

\subsection*{Hierarchical clustering of ECoG features}

\begin{figure}[t]
\centering
\includegraphics[width=0.8\linewidth]{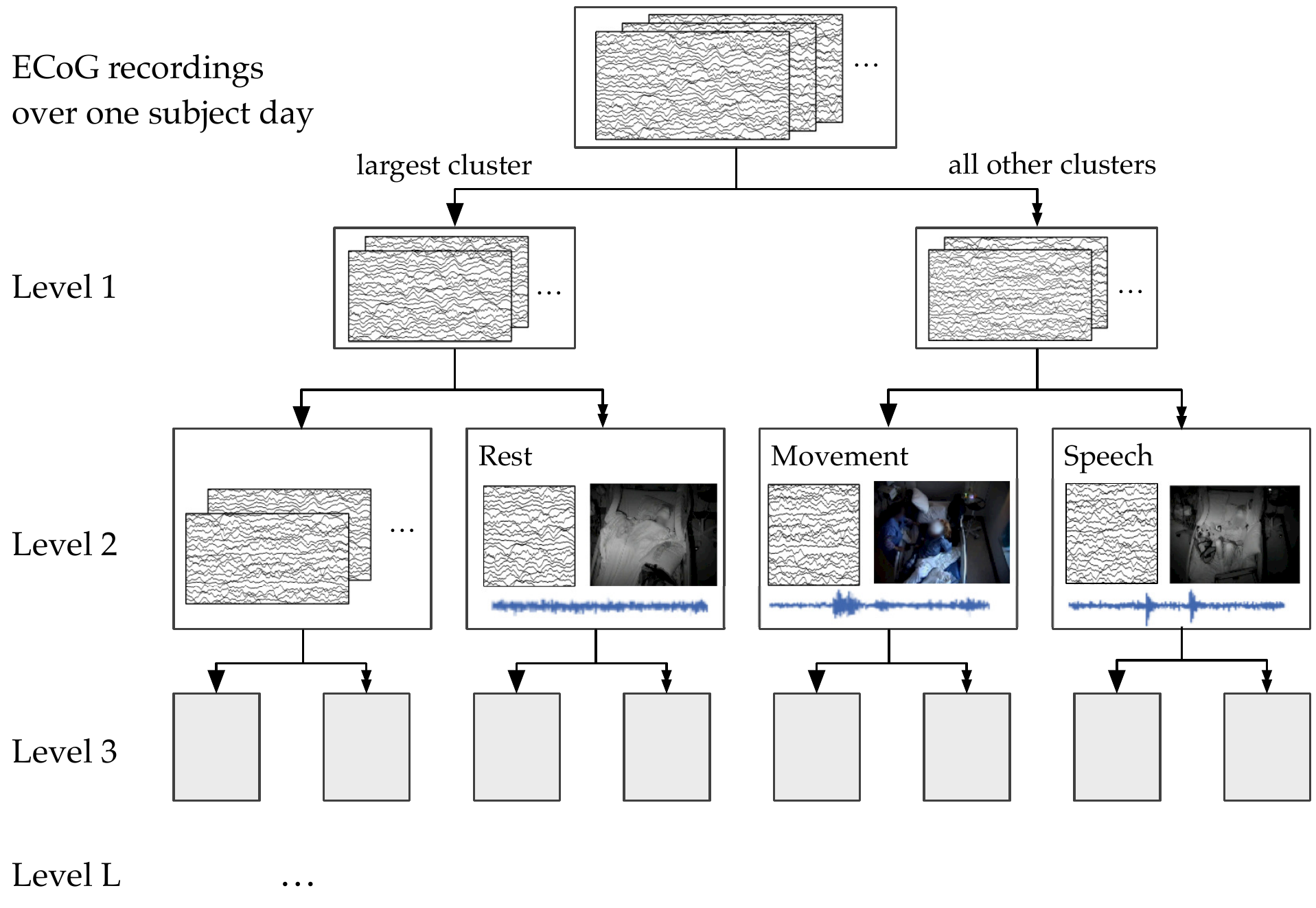}
\caption{A schematic of the hierarchical clustering and annotation method. Features extracted from ECoG recordings of each subject day were recursively clustered and agglomerated at increasing levels. Annotation consisted of finding the cluster within each level whose time course had the highest correlation with automatically extracted movement and speech levels. For illustration purposes, here we show the annotation of clusters at Level 2, which has 4 total clusters.}
\label{fig:hierarchy}
\end{figure}

We used the 50-dimensional principal component power spectral features of the ECoG data as features for our hierarchical k-means clustering. 
The hierarchical k-means procedure and the annotation of clusters by correlation with movement and speech levels is shown schematically in Fig.~\ref{fig:hierarchy}.
For each subject on each day, we first perform k-means with $k=20$ clusters.
Next, we segregate the data points into the single cluster with the most number of data points and all the rest of the clusters.
This procedure produces the first level of the hierarchical clustering, which now has two clusters.
Next, for level $L$ of the clustering, this procedure is repeated for each cluster from level $L-1$ using $k = 20/L$ ($k$ floored to the largest previous integer).
Again, the single cluster with the most number of data points is separated from the rest of the clusters, so that at level $L$, we end up with $2^L$ clusters.
This process of recursive k-means clustering and aggregation is stopped when there are fewer than 100 data points in each cluster, or when $L=10$.
In this manuscript, we focused on analyzing annotated clusters in levels 1--4.

\subsection*{Automated annotation of clusters using video and audio recordings}

Results from the clustering analysis were automatically annotated using movement and speech levels.
Each type of unsupervised analysis produced time series at different temporal resolutions, so they were all first consolidated into a mean analog value for non-overlapping 16-second windows.
For ECoG, we counted how many 2-second windows within each 16-seconds were assigned to a particular cluster at the target hierarchy level.
For movement and speech detection, we considered what fraction of the 16-second window exceeded a threshold value. 
These thresholds were determined empirically.

After consolidation into windows of 16 seconds, we computed the Pearson r correlation between each of the ECoG clusters with the movement and speech levels.
The ``movement'' and ``speech'' labels were assigned to clusters for which the correlation was the highest for each behavior.
If movement and speech both correlated best with the same cluster, the label was assigned to the second best cluster for the activity type that had a lower correlation.
The ``rest'' label was assigned to the cluster with the largest negative correlation with both movement and speech.

We performed this annotation assignment for clustering levels 1--4 (see Fig.~\ref{fig:hierarchy}).
At level 1, for which there were only 2 total clusters, labels were simplified to be ``rest'' and ``non-rest'' (for movement and speech combined).
Level 3, where there are 8 total clusters, appeared to be the most parsimonious level of granularity for the number of categories available automatically.

\subsection*{Validation with ground truth labels}

Ground truth labels for random portions of each day were obtained from two students who hand annotated a small random fraction of each subject day (about 40 minutes, or 3\% of each day) for visible and audible behaviors. The hand annotations were distributed randomly throughout each day of each patient.
The automated results were compared to manual labels using 16 second windows within the manually annotated times. 
Each 16 second window is determined to contain an activity if the activity is annotated within any point in the window. 
Since different clusters have different baseline levels, we determined that the cluster detects the activity if its level is at or above the 25th percentile over the day. 
Using the manual labels as ground truth, accuracy and F1 scores were computed.
The F1 score is the harmonic mean of precision and recall.

To determine the statistical significance of the F1 scores compared to chance, we generated shuffled labels by changing the timing of the ground truth labels of each activity, without changing their overall relative frequency.
This shuffling was repeated over 1000 random iterations to determine the distribution of F1 scores assuming chance, and the true F1 score was compared against these shuffled F1 scores.
We report the percentile of the true F1 scores for all subject days.

\subsection*{Mapping annotated clusters back to the brain}

For each annotated cluster, we projected the centroid values of the cluster back to brain coordinates.
The centroids values are 50-dimensional vectors in PCA space, reduced from power spectral features of all recording electrodes.
The inverse PCA transform using the original PCA basis projects the centroid back to brain coordinates, where the relative power in each frequency bin is available at each electrode.
Note that because of the Z-Score normalization step before computing the original PCA basis, this back-projection reproduces Z-Score values, not voltages.
These Z-Scores can be separately averaged according to frequency bins of interesting bands, including a low frequency band (LFB, 1--8 Hz) and a high-frequency band (HBG, 12--45 Hz) as show in Fig.~\ref{fig:backprojection}.
Fig. S5--S8 also show results of brain maps at 72--100Hz.
Anatomic reconstruction of electrode coordinates on structural imaging of subjects' brains was available for only four of the six subjects, so we were unable to perform this mapping for Subject 2 and Subject 6.

\section*{Acknowledgments}
We thank Ryan Shean and Sharon Ke for annotating the videos and audio recordings for ground truth labels. We also thank the staff and doctors at Harborview Medical Center, in particular Julie Rae, Shahin Hakimian and Jeffery Tsai for aiding in data collection as well as research discussions. James Wu contributed to brain reconstructions. We thank Ali Farhadi for giving advice on the computer vision based video processing and Lise Johnson for discussions about unsupervised clustering.

\textbf{Funding.} This research was support by the Washington Research Foundation (WRF), the National Institutes of Health (NIH) grants NS065186, 2K12HD001097-16 and 5U10NS086525-03, and award EEC-1028725 from the National Science Foundation (NSF).

\section*{Author contributions statement}
N.W. conceived the experiments with the help of J.D.O., J.G.O.; N.W. and R.R. conducted the experiment; N.W., R.R., and B.B. performed the analysis and wrote the manuscript, with input from all authors.

\section*{Additional information}
The authors declare no competing financial interests.
\end{document}